\documentclass[aps,pre,reprint,showpacs,superscriptaddress,nofootinbib]{revtex4-2}

\usepackage{graphicx}
\usepackage{dcolumn}
\usepackage{bm}
\usepackage{xcolor}  
\definecolor{changecol}{rgb}{0.0,0.00,0.00}
\expandafter\let\csname equation*\endcsname\relax
\expandafter\let\csname endequation*\endcsname\relax
\usepackage{amsmath}
\usepackage{graphicx}
\usepackage{soul}
\usepackage{float}
\usepackage{physics}
\usepackage{comment}
\emergencystretch=\maxdimen
\usepackage[export]{adjustbox}
\usepackage{hyperref}
\hypersetup{
    pdfborder={0 0 0} 
}
\usepackage{orcidlink}

\begin{document}

\title{Pointwise mutual information bounded by stochastic Fisher information}

\author{Pedro \surname{B.~Melo}\,\orcidlink{0000-0002-3726-761X}}
\email{pedrobmelo@aluno.puc-rio.br}
 \affiliation{Departamento de F\'isica, PUC-Rio, 22452-970, Rio de Janeiro RJ, Brazil}
 
\date{\today}

\begin{abstract}
{\color{changecol}How much information about an unknown parameter can be assigned to one realized outcome, and how is that information constrained by the sensitivity of the same outcome? We derive upper bounds on the pointwise mutual information in terms of the stochastic Fisher information, both for parameters in a bounded interval and for sufficiently regular differentiable priors, and state the corresponding saturation conditions. Averaging the pointwise inequality and applying Jensen's inequality twice yields the known mutual-information--Fisher-information bound, while making explicit the outcome dependence discarded by the ensemble description. We illustrate the classical result using the instantaneous position of an overdamped Langevin particle as the observed outcome. For a fixed, parameter-independent quantum measurement, the stochastic Fisher information is further bounded by the conditional quantum Fisher information (CQFI), whose probability-weighted mean is the quantum Fisher information. A thermal-qubit field-sensing example evaluates this finite-prior bound deterministically from exact $2\times2$ matrices and, separately, decomposes the projected SLD second moment into population, coherent, and interference contributions. The interference contribution can be negative for an individual probe direction and has zero unweighted trace, but it is not itself an operational measure of readout quality. Neither the CQFI nor its mean selects a measurement basis: the latter is measurement-independent, whereas the relevant basis-dependent local objective is the classical Fisher information of the chosen measurement.}
\end{abstract}

\maketitle

\section{INTRODUCTION}

{\color{changecol}The Fisher information (FI) quantifies the local statistical sensitivity of a likelihood to an unknown parameter and, through the Cram\'er--Rao inequality, bounds the variance of regular unbiased estimators. Its quantum extension is obtained by optimizing the classical FI over measurements \cite{BraunsteinCaves1994,Paris2009,DemkowiczDobrzanski_2020,Liu_2020} and underpins sensing in optical clocks, gravitational-wave interferometers, and magnetometry \cite{Ludlow2015,Tse2019,Degen2017}. It has also become experimentally accessible: randomized-measurement and variational protocols can estimate the QFI or controlled bounds on it on present-day platforms \cite{Rath2021,Beckey2022}.}

{\color{changecol}The mutual information (MI) provides the corresponding global Bayesian quantity: it is the expected log-likelihood ratio, or expected information gain, between parameter and outcome \cite{CoverThomas2006}. It is therefore a natural objective in Bayesian and adaptive metrology \cite{HentschelSanders2011,Fiderer2021}, in communication-theoretic analyses of sensing, and in neural coding, where the relation between FI and MI has been studied for decades \cite{brunel1998mutual,wei2016mutual,Lu2024}. With the natural logarithms used below, information is measured in nats. G\'orecki \textit{et al.} \cite{Gorecki_2025} recently established a general upper bound on MI in terms of FI and the prior, valid for broad prior families and free of the divergences that affect Bayesian Cram\'er--Rao bounds for sharp priors \cite{van2004detection,ziv1969some}.}

{\color{changecol}Ensemble quantities remain the appropriate benchmarks for repeated and asymptotic estimation, but they do not determine the information carried by a particular record. Outcome-resolved descriptions become useful when an inference or control decision must be conditioned on the current readout, as in single-molecule nonequilibrium measurements \cite{Collin2005}, single-shot qubit readout \cite{Walter2017}, and continuously monitored systems whose state and feedback depend on the measurement record \cite{GammelmarkMolmer2013,Murch2013,Vijay2012}. They are also useful when rare or non-stationary realizations matter \cite{Fogelmark2018,Radaelli_2023}. Related record-level descriptions are central in stochastic and information thermodynamics \cite{Seifert2012,Parrondo2015} and in continuously monitored quantum thermodynamics \cite{elouard2019work,manzano2022quantum,ferri2025conditional}. Importantly, an outcome-resolved analysis still presupposes a calibrated likelihood and prior; one observation alone does not determine $p(x|\theta)$ or $p(x)$. Another important application is syndrome-based noise tracking in quantum error correction is one prospective application. Syndrome statistics can reveal physical noise parameters without measuring the encoded logical state \cite{Wagner2021,Wagner2022}, and error-detection events have recently been used as learning signals for compensating processor drift \cite{Sivak2026}.}

{\color{changecol}The corresponding outcome-resolved quantities are known. The information density, also called the pointwise mutual information (PMI), is the log-likelihood ratio assigned to one parameter--outcome pair and averages to the MI \cite{CoverThomas2006,makkeh2021introducing,williams2022suspicious,czyz2024propertiesestimationpointwisemutual}. The squared score evaluated on one outcome is here called the stochastic Fisher information (SFI) \cite{Melo_STFI,melo2025}; it averages to the FI and has been used to formulate speed limits for individual noisy realizations. An outcome $x$ may itself be an entire time record, but the theorems do not require this: they apply equally to a single discrete outcome or instantaneous readout. What has been missing is an outcome-resolved analogue of the MI--FI bound of Ref.~\cite{Gorecki_2025}.}

{\color{changecol}In this work we supply it. We derive PMI bounds directly from the SFI for a bounded parameter interval and for sufficiently regular differentiable priors (Theorems~1 and~2). Averaging the pointwise inequality and applying Jensen's inequality twice yields the bound of Ref.~\cite{Gorecki_2025}; thus the pointwise result retains outcome dependence discarded by the two ensemble relaxations. We then map the conditional likelihood onto the Born rule for a fixed, parameter-independent POVM and obtain a quantum bound in terms of the conditional quantum Fisher information (CQFI) introduced in Ref.~\cite{melo2026} (Theorem~3).}

{\color{changecol}The quantum case requires care. The CQFI is defined through the symmetric logarithmic derivative and conditioned on an outcome; its probability-weighted mean is the QFI. A related but distinct quantity, the projected SLD second moment, admits a decomposition into population, coherent, and interference contributions. The interference contribution can be negative for one probe direction, and its unweighted sum over a complete orthonormal basis vanishes; this is not a probability-weighted average, which need not vanish \cite{melo2026}. We compute the three projected-SLD contributions analytically for a thermal qubit, but do not interpret the cross-term as a basis-selection criterion.}

{\color{changecol}This work is organized as follows. Section~\ref{sec:bounds} introduces the PMI and SFI and states the classical bounds and their achievability conditions. Section~\ref{sec:average_bounds} derives the ensemble MI--FI bound from the pointwise inequality. Section~\ref{sec:quantum} establishes the quantum generalization and distinguishes the CQFI from the projected SLD second moment. Section~\ref{sec:examples} treats an overdamped Langevin readout and thermal-qubit field sensing. Section~\ref{sec:conclusions} summarizes the scope and limitations, including syndrome-based noise tracking as an outlook. Appendices~\ref{app:A} and~\ref{app:B} contain the proofs.}

\section{BOUNDS ON PMI}
\label{sec:bounds}

We first define the PMI and SFI together with their ensemble counterparts. We then state the pointwise bounds, derive them from a common total-variation inequality, and discuss their scope and achievability.

\subsection{Definitions and their ensemble counterparts}

Consider an unknown parameter $\theta$ drawn from a prior distribution $p(\theta)$, and a measurement outcome $x$ governed by a conditional probability $p(x|\theta)$. The local information gained from a single realization is quantified by the PMI \cite{czyz2024propertiesestimationpointwisemutual}, which compares the joint probability to the product of marginals
\begin{equation}
i(x,\theta)=\log\frac{p(x,\theta)}{p(x)p(\theta)} := \log{\frac{p(x|\theta)}{p(x)}},
\label{eq:PMI}
\end{equation}
where the last equality follows from Bayes' rule $p(x|\theta) = p(x,\theta)/p(\theta)$.

The PMI is the kernel of the mutual information. Averaging Eq.~\eqref{eq:PMI} over the joint distribution recovers the standard definition,
\begin{equation}
I(X,\Theta) = \sum_{x}\int d\theta \, p(x,\theta) \log\frac{p(x,\theta)}{p(x)p(\theta)} = \langle i(x,\theta)\rangle,
\label{eq:MI}
\end{equation}
\begingroup
\color{changecol}
so that every statement about the PMI concerns the integrand of the MI before averaging. The MI is non-negative because it is a Kullback--Leibler divergence, whereas the PMI need not be. One has $i(x,\theta)<0$ whenever $p(x|\theta)<p(x)$: the outcome supplies negative local evidence for that parameter value relative to the prior-predictive distribution. This does not by itself mean that an estimator points away from the true parameter. The outcome dependence, including these negative contributions, is information that the ensemble average does not retain.

The local sensitivity of the realized outcome to the parameter is captured by the stochastic Fisher information (SFI) \cite{Melo_STFI,melo2025},
\begin{equation}
\iota(x,\theta)=\left(\frac{\partial\log p(x|\theta)}{\partial\theta}\right)^{2}.
\label{eq:SFI}
\end{equation}
The SFI stands to the FI exactly as the PMI stands to the MI. The classical Fisher information is defined as the expectation of the squared score, $F(\theta) = \sum_x p(x|\theta)\left(\partial_\theta \log p(x|\theta)\right)^2$; comparing with Eq.~\eqref{eq:SFI},
\begin{equation}
F(\theta) = \sum_{x} p(x|\theta)\, \iota(x,\theta) = \langle \iota(x,\theta)\rangle_{x|\theta},
\label{eq:FI_from_SFI}
\end{equation}
so the SFI is the kernel of the FI evaluated on a single realization, and the FI is recovered by averaging over outcomes at fixed $\theta$. Operationally, $-\log p(x|\theta)$ is the surprisal of the outcome; its sensitivity to the parameter is the score $\partial_\theta \log p(x|\theta)$, and the SFI is the square of that score. Unlike $F(\theta)$, which is a deterministic function of $\theta$ alone, $\iota(x,\theta)$ fluctuates from run to run, and its variance is a genuine property of the process that the FI does not record.
\endgroup

\subsection{Classical bounds}

Building on Eqs.~\eqref{eq:PMI} and \eqref{eq:SFI}, we now establish upper bounds on the PMI that depend directly on the local fluctuations of the system. It has been shown in Ref.~\cite{Gorecki_2025} that the average mutual information obeys a bound given by the Fisher information and the prior $p(\theta)$. The following two theorems generalize that result to the trajectory level.

\begingroup
\color{changecol}
\textit{Theorem 1.} If the support of the unknown parameter lies in a bounded interval, $\operatorname{supp}p(\theta)\subseteq[a,b]$, then the PMI is bounded by
\begin{align}
i(x,\theta) &\le \log\left(\frac{p(x|a)+p(x|b)}{2p(x)} \nonumber \right. \\ &\left.+ \frac{1}{2}\int_{a}^{b}d\theta\left|\frac{p(x|\theta)}{p(x)}\right|\sqrt{\Lambda_{1}(x,\theta)}\right)\label{eq:THM1}
\end{align}
where $\Lambda_{1}(x,\theta)=\iota(x,\theta)$.

\begingroup
\color{changecol}
For fixed $x$, define the total variation of the likelihood on $[a,b]$ by
\begin{equation}
\operatorname{TV}_{[a,b]}\!\bigl(p(x|\cdot)\bigr)
:=\int_a^bdu\,\left|\partial_up(x|u)\right|.
\label{eq:TV_definition}
\end{equation}

\textit{Remark 1 (achievability of Theorem 1).} The bound is saturated when two conditions hold. First, the evaluation point $\theta$ must be a global maximum-likelihood estimate within $[a,b]$ for the observed outcome $x$. Second, the total variation of the likelihood must obey
\begin{equation}
\operatorname{TV}_{[a,b]}\!\bigl(p(x|\cdot)\bigr)
=2\max_{u\in[a,b]}p(x|u)-p(x|a)-p(x|b).
\label{eq:TV_saturation}
\end{equation}
This identity holds for every monotone likelihood and for every likelihood that rises monotonically to a single maximum and then falls monotonically. Additional oscillations introduce slack. If both endpoint likelihoods vanish, the right-hand side reduces to twice the maximum, but endpoint vanishing alone is not sufficient in the presence of additional oscillations. Away from saturation the bound remains valid, and its gap can be evaluated outcome by outcome, as illustrated in Sec.~\ref{sec:langevin}.
\endgroup

\textit{Theorem 2.} Fix an outcome $x$ with $p(x)>0$. Let $p(\theta)$ be differentiable and suppose that $q_x(\theta):=p(x|\theta)p(\theta)$ has finite total variation and satisfies $q_x(\theta)\to0$ as $\theta\to\pm\infty$. Wherever the score is used we assume $p(x|\theta)>0$, or interpret it by a continuous limiting prescription. Then
\begin{equation}
i(x,\theta) \le \log\left(\frac{1}{2}\int_{-\infty}^{\infty}d\theta\left|\frac{p(x|\theta)}{p(x)}\right|\sqrt{\Lambda_{2}(x,\theta)}\right)+s(\theta)
\label{eq:THM2}\end{equation}
where
\begin{align}
\Lambda_{2}(x,\theta)&=\iota(x,\theta)p(\theta)^{2}+\dot{p}(\theta)^{2} + 2\partial_{\theta}\log p(x|\theta)\dot{p}(\theta)p(\theta).
\label{eq:Lambda2}
\end{align}
Here, $\dot{p}(\theta)=dp(\theta)/d\theta$ and $s(\theta)=-\log p(\theta)$ is the surprisal of the prior. It coincides with a stochastic entropy only when $\theta$ itself labels the relevant physical random state. The bound is not parameter invariant, highlighting its sensitivity to the chosen parametrization.

\textit{Remark 2 (achievability of Theorem 2).} Saturation requires that (i) the evaluation point $\theta$ be a global maximizer of $p(x|\theta)p(\theta)$ and (ii) this product rise monotonically to that maximum and fall monotonically thereafter. The required boundary behavior is already part of Theorem~2. Condition (ii) is a joint statement about likelihood and prior: a unimodal likelihood combined with a multimodal prior need not saturate the bound.
\endgroup

\subsection{Derivation}

To derive Eqs.~\eqref{eq:THM1} and \eqref{eq:THM2}, we introduce an arbitrary, non-negative function $f(\theta)$ such that $\text{supp} f(\theta) \supseteq \text{supp} p(\theta)$, and rewrite the PMI algebraically as
\begin{equation}
    i(x,\theta) = \log\frac{p(x|\theta)f(\theta)}{p(x)} - \log f(\theta).
    \label{eq:pmi_rewrite}
\end{equation}
{\color{changecol}To bound the argument of the first logarithm, we use total variation. We assume that $p(x|\theta)f(\theta)\to0$ as $\theta\to\pm\infty$ and that this product has finite total variation. The quantity at a specific evaluation point is bounded by its global maximum, which the fundamental theorem of calculus bounds in turn by half the integral of the absolute derivative,}
\begin{align}
    \frac{p(x|\theta)f(\theta)}{p(x)} &\le \max_{\theta'} \left( \frac{p(x|\theta')f(\theta')}{p(x)} \right) \nonumber \\&\le \frac{1}{2}\int_{-\infty}^{\infty} d\theta' \left| \frac{d}{d\theta'} \left( \frac{p(x|\theta')f(\theta')}{p(x)} \right) \right|.
    \label{eq:maxbound}
\end{align}
Expanding the derivative inside the absolute value using the product rule, and recognizing that probability distributions are strictly non-negative, we factor out $p(x|\theta')/p(x)$. Using the identity $\partial_\theta p(x|\theta) = p(x|\theta)\partial_{\theta}\log p(x|\theta)$, the remaining terms are collected inside a square root, which naturally introduces the SFI of Eq.~\eqref{eq:SFI}. Defining the general functional
\begin{align}
    \Lambda(x,\theta) &\equiv \iota(x,\theta)f(\theta)^2 + \dot{f}(\theta)^2 \nonumber\\&+ 2\partial_\theta \log p(x|\theta) f(\theta)\dot{f}(\theta),
    \label{eq:Lambda_gen}
\end{align}
and substituting back into Eq.~\eqref{eq:pmi_rewrite}, we obtain the generalized bound
\begin{equation}
    i(x,\theta) \le \log \left( \frac{1}{2}\int_{-\infty}^{\infty} d\theta' \frac{p(x|\theta')}{p(x)} \sqrt{\Lambda(x,\theta')} \right) - \log f(\theta).\label{eq:gen_bound}
\end{equation}

Both theorems follow from Eq.~\eqref{eq:gen_bound} through specific choices of $f(\theta)$.

For Theorem 1, with $\text{supp}~ p(\theta) \subseteq [a,b]$, we choose $f(\theta)$ to be a boxcar function over $[a,b]$. The subadditivity property $\sqrt{y+z} \le \sqrt{y} + \sqrt{z}$ decouples the SFI from the absolute derivative $|\dot{f}(\theta)|$, avoiding the square of a step-function derivative. The latter yields Dirac delta functions at the boundaries which, integrated against the $1/2$ factor, extract the halved boundary terms $p(x|a)/2p(x)$ and $p(x|b)/2p(x)$. Within the open interval $(a,b)$ one has $f(\theta)=1$ and $\dot{f}(\theta)=0$, so that $\Lambda_1(x,\theta) = \iota(x,\theta)$, and the penalty term $-\log f(\theta)$ vanishes.

For Theorem 2, we set $f(\theta) = p(\theta)$. Substitution into Eq.~\eqref{eq:Lambda_gen} immediately recovers $\Lambda_2(x,\theta)$ of Eq.~\eqref{eq:Lambda2}, and the penalty term $-\log f(\theta)$ becomes $-\log p(\theta) = s(\theta)$ [cf. Appendix \ref{app:A}].

\subsection{Discussion}

{\color{changecol}These outcome-resolved bounds provide information not contained in an ensemble benchmark: they constrain the particular readout on which an inference or feedback decision is conditioned. Ensemble averages remain accessible and useful whenever repeated calibration data or a likelihood model are available, but they do not determine the fluctuation associated with the realized outcome. In common with their ensemble counterparts \cite{Gorecki_2025}, the present bounds also avoid the divergences found in traditional Bayesian Cram\'er--Rao limits for sharp priors \cite{van2004detection,ziv1969some}, such as a uniform distribution over a finite interval.}

\section{RECOVERING THE AVERAGE BOUNDS}
\label{sec:average_bounds}
{\color{changecol}Having established the outcome-dependent PMI bounds, we now relate them to the global MI bound of Ref.~\cite{Gorecki_2025}. Averaging the pointwise inequality is followed by two applications of Jensen's inequality. These relaxations discard outcome dependence and yield the known ensemble functional form; they do not imply that the ensemble bound is literally the probability average of the pointwise right-hand side.}

The MI is defined as the ensemble average of the PMI over the joint probability distribution $p(x,\theta) = p(x|\theta)p(\theta)$, such that $I(X,\Theta) = \langle i(x,\theta) \rangle$. To find the average bound, we take the expectation value of the generalized bound from Eq. (\ref{eq:gen_bound}).

Averaging over the joint distribution yields
\begin{align}
    I(X,\Theta) &\le \sum_x \int d\theta~p(x,\theta) \nonumber\\&\times\left[ \log \left( \frac{1}{2}\int d\theta' \left| \frac{p(x|\theta')}{p(x)} \right| \sqrt{\Lambda(x,\theta')} \right) - \log f(\theta) \right] \nonumber \\
    &= \sum_x p(x) \log \left( \frac{1}{2}\int d\theta' \frac{p(x|\theta')}{p(x)} \sqrt{\Lambda(x,\theta')} \right) \nonumber\\&- \int d\theta p(\theta) \log f(\theta).
\end{align}
Notice that the second term immediately recovers the penalty term associated with the arbitrary function $f(\theta)$ found in standard MI bounds. 

To evaluate the first term, we utilize Jensen's inequality. Since the logarithm is a strictly concave function, the average of the logarithm is bounded by the logarithm of the average: $\sum_x p(x) \log(A_x) \le \log(\sum_x p(x) A_x)$. Applying this to the sum over $x$, we obtain
\begin{align}
    &\sum_x p(x) \log \left( \frac{1}{2}\int d\theta' \frac{p(x|\theta')}{p(x)} \sqrt{\Lambda(x,\theta')} \right) \nonumber\\&\le \log \left( \sum_x p(x) \frac{1}{2}\int d\theta' \frac{p(x|\theta')}{p(x)} \sqrt{\Lambda(x,\theta')} \right) \nonumber \\
    &= \log \left( \frac{1}{2}\int d\theta' \sum_x p(x|\theta') \sqrt{\Lambda(x,\theta')} \right).
\end{align}

We apply Jensen's inequality a second time to the sum inside the integral, $\sum_x p(x|\theta') \sqrt{\Lambda(x,\theta')} \le \sqrt{ \sum_x p(x|\theta') \Lambda(x,\theta') }$. We now evaluate the expectation value of our functional $\Lambda(x,\theta)$ with respect to the conditional probability
\begin{align}
    \sum_x p(x|\theta) \Lambda(x,\theta) &= \sum_x p(x|\theta)\left[ \iota(x,\theta)f(\theta)^2 + \dot{f}(\theta)^2 \right.\nonumber\\&\left.+ 2\partial_\theta \log p(x|\theta) f(\theta)\dot{f}(\theta) \right].
\end{align}

By definition, the average of the stochastic Fisher information (SFI) is the classical Fisher information (FI), $\sum_x p(x|\theta)\iota(x,\theta) = F(\theta)$. Furthermore, the cross-term vanishes because the sum of the derivative of a probability distribution is zero
\begin{align}
    \sum_x p(x|\theta) \partial_\theta \log p(x|\theta) &= \sum_x \frac{\partial p(x|\theta)}{\partial \theta} =  0.
\end{align}

Consequently, the expectation of the functional simplifies purely to ensemble quantities
\begin{equation}
    \sum_x p(x|\theta) \Lambda(x,\theta) = F(\theta)f(\theta)^2 + \dot{f}(\theta)^2.
\end{equation}

Substituting this back into our averaged bound, we arrive at
\begin{align}
    I(X,\Theta) &\le \log \left( \frac{1}{2}\int d\theta \sqrt{F(\theta)f(\theta)^2 + \dot{f}(\theta)^2} \right) \nonumber\\&- \int d\theta p(\theta) \log f(\theta).
\end{align}

{\color{changecol}This recovers the functional form of the mutual-information bound of Ref.~\cite{Gorecki_2025}. Thus the known ensemble result follows as a corollary of the pointwise inequality together with the two Jensen relaxations, while the pointwise statement retains the dependence on the realized outcome.}

\section{QUANTUM GENERALIZATION}
\label{sec:quantum}

\begingroup
\color{changecol}
To extend the outcome-resolved bounds to quantum systems, we map classical conditional probabilities onto quantum measurements. The likelihood is determined by the Born rule, $p(x|\theta)=\text{Tr}(\Pi_x\rho_\theta)$, where $\{\Pi_x\}$ is a fixed positive operator-valued measure (POVM), independent of the unknown parameter over the domain of the bound, and $\rho_\theta$ is the parameter-dependent density matrix. This is not a Holevo-type accessible-information bound: the Holevo quantity constrains the measurement-optimized ensemble information of a classical--quantum ensemble, whereas the present result fixes the measurement and bounds the information density of each resulting classical outcome through its local parameter sensitivity \cite{Holevo1973,BraunsteinCaves1994}.

\subsection{The conditional quantum Fisher information}

Mapping the classical SFI $\iota(x,\theta)$ to an outcome-resolved quantum quantity requires care. The conditional quantum Fisher information (CQFI) was introduced in Ref.~\cite{melo2026}. For a specific measurement outcome $x$, it is defined through the symmetric logarithmic derivative (SLD) operator $L_\theta$, the Hermitian solution of the Lyapunov equation $\partial_\theta \rho_\theta = \frac{1}{2}\{L_\theta,\rho_\theta\}$, as
\begin{equation}
f_{Q,x}(\theta)=\frac{\text{Tr}(\Pi_{x}L_{\theta}\rho_{\theta}L_{\theta})}{\text{Tr}(\rho_{\theta}\Pi_{x})}.
\label{eq:CQFI}
\end{equation}
The symmetric ordering $L_\theta \rho_\theta L_\theta$ is produced by the Cauchy--Schwarz [cf. Appendix~\ref{app:B}] and guarantees that $f_{Q,x}(\theta)$ is real and non-negative. The superficially similar expression $\text{Tr}(\Pi_x L_\theta^2 \rho_\theta)/\text{Tr}(\rho_\theta \Pi_x)$ can be complex for a rank-one $\Pi_x$ that is not an eigenprojector of $\rho_\theta$; commutation of $\rho_\theta$ and $L_\theta$ is a sufficient condition for the two operator orderings to agree \cite{melo2026}. The CQFI fluctuates from outcome to outcome and satisfies $\sum_xp(x|\theta)f_{Q,x}(\theta)=\mathcal F_Q(\theta)$ by completeness.

Applying the Cauchy-Schwarz inequality to the SLD for a fixed outcome yields [cf. Appendix \ref{app:B}]
\begin{equation}
\iota(x,\theta) \le f_{Q,x}(\theta),
\label{eq:sfi_le_cqfi}
\end{equation}
which is the inequality that carries the classical theorems into the quantum domain.
\endgroup

\begingroup
\color{changecol}
\subsection{The projected SLD second moment and its decomposition}

A second, closely related object is useful for physical interpretation. For a rank-one projector $\Pi_\alpha = |\alpha\rangle\langle\alpha|$, Ref.~\cite{melo2026} introduces the \textit{projected SLD second moment}
\begin{equation}
\tilde{f}_{Q,\alpha}(\theta) = \text{Tr}(\Pi_\alpha L_\theta^2) = \langle\alpha|L_\theta^2|\alpha\rangle,
\label{eq:proj_SLD}
\end{equation}
which measures the local power of the SLD generator along a chosen direction in Hilbert space. We emphasize that $\tilde{f}_{Q,\alpha}$ and $f_{Q,x}$ are distinct quantities: writing $\rho_\theta=\sum_np_n|n\rangle\langle n|$ and $c_n=\langle n|\alpha\rangle$, Eq.~\eqref{eq:CQFI} is a $p_m$-weighted sum divided by the outcome probability, whereas Eq.~\eqref{eq:proj_SLD} contains the corresponding unweighted sum. Their exact equality condition is the weighted cancellation
\begin{equation}
\sum_m\left[p_m-\langle\alpha|\rho_\theta|\alpha\rangle\right]
\left|\langle\alpha|L_\theta|m\rangle\right|^2=0.
\label{eq:equality_cond}
\end{equation}
Requiring termwise equality of the spectral weights is sufficient but not necessary. Equation~\eqref{eq:equality_cond} also holds whenever $|\alpha\rangle$ is an eigenvector of $L_\theta$, because both quantities then equal the corresponding squared SLD eigenvalue.

It is $\tilde{f}_{Q,\alpha}$, not $f_{Q,x}$, that admits the three-way spectral decomposition \cite{melo2026}
\begin{equation}
\tilde{f}_{Q,\alpha}(\theta) = \tilde{f}_{Q,\alpha}^{IC}(\theta) + \tilde{f}_{Q,\alpha}^{C}(\theta) + \tilde{f}_{Q,\alpha}^{X}(\theta),
\label{eq:decomp}
\end{equation}
into an incoherent term arising from population changes, a coherent term arising from rotations of the eigenbasis, and a cross-term interfering the two [cf. Appendix~\ref{app:B}]. The incoherent and coherent terms are non-negative; the cross-term $\tilde{f}^{X}_{Q,\alpha}$ can be negative, signalling destructive interference between population reweighting and basis rotation relative to the probe direction. Its unweighted sum over a complete orthonormal basis is zero because $\Tr\{L_{IC},L_C\}=0$, but its probability-weighted average is generally non-zero.

The distinction matters for what follows. The quantity appearing in our bounds is $f_{Q,x}$, whereas the interference mechanism is a property of $\tilde{f}_{Q,x}$. The relation between the two is not an inequality of fixed direction. Comparing the weighted sum \eqref{eq:CQFI} with the unweighted sum \eqref{eq:proj_SLD} in the eigenbasis of $\rho_\theta$ [cf. Appendix \ref{app:B}] gives the exact criterion
\begin{equation}
f_{Q,x}(\theta) \le \tilde{f}_{Q,x}(\theta) \iff \sum_m \left(p_m - \langle x|\rho_\theta|x\rangle\right)\left|\langle x|L_\theta|m\rangle\right|^2 \le 0,
\label{eq:chain_cond}
\end{equation}
a weighted statement whose sign depends on the probe direction and spectrum. Both signs occur in the thermal-qubit model of Sec.~\ref{sec:qubit}; neither ordering is universal, and no general chain $\iota\le f_Q\le\tilde f_Q$ is available.

Consequently we state Theorem 3 below in terms of $f_{Q,x}$ throughout, which is the quantity the Cauchy-Schwarz step actually delivers, and we treat the decomposition \eqref{eq:decomp} as diagnosing the physical content of the trajectory's sensitivity rather than as a term-by-term account of the bound. Where the equality condition \eqref{eq:equality_cond} is met --- for example for an SLD eigenvector, or for a shared eigenprojector in the commuting case $[\rho_\theta,L_\theta]=0$ --- the two objects coincide and the decomposition applies to the bound directly.
\endgroup

\subsection{Quantum bounds}
\begingroup
\color{changecol}

By applying the mapping $\iota(x,\theta) \rightarrow f_{Q,x}(\theta)$ justified by Eq.~\eqref{eq:sfi_le_cqfi}, the structure of the classical theorems extends directly to single-shot quantum measurements.

\textit{Theorem 3.} Let $\rho_\theta$ be a differentiable family of quantum states with prior $p(\theta)$, and let a fixed POVM $\{\Pi_x\}$, independent of $\theta$, produce the classical outcome $x$ with $p(x|\theta)=\text{Tr}(\Pi_x\rho_\theta)$.

If $\operatorname{supp}p(\theta)\subseteq[a,b]$, the PMI of the classical outcome generated by this quantum measurement is bounded by
\begin{align}
i(x,\theta) \le& \log\left(\frac{p(x|a)+p(x|b)}{2p(x)}   \right.\nonumber\\
&\left. +\frac{1}{2}\int_{a}^{b}d\theta\left|\frac{p(x|\theta)}{p(x)}\right|\sqrt{\Lambda_{1,Q}(x,\theta)}\right),
\label{eq:THM3a}
\end{align}
where $\Lambda_{1,Q}(x,\theta)=f_{Q,x}(\theta)$.

For an infinite-support prior satisfying the regularity and boundary assumptions of Theorem~2, the same classical-outcome PMI is bounded by
\begin{equation}
i(x,\theta) \le \log\left(\frac{1}{2}\int_{-\infty}^{\infty}d\theta\left|\frac{p(x|\theta)}{p(x)}\right|\sqrt{\Lambda_{2,Q}(x,\theta)}\right)+s(\theta),
\label{eq:THM3b}
\end{equation}
where $\Lambda_{2,Q}(x,\theta)=f_{Q,x}(\theta)p(\theta)^{2}+\dot{p}(\theta)^{2}+2\partial_{\theta}\log p(x|\theta)\dot{p}(\theta)p(\theta)$ and $s(\theta)=-\log p(\theta)$.

\textit{Proof.} Since the POVM is fixed, $\partial_\theta p(x|\theta)=\text{Tr}(\Pi_x\partial_\theta\rho_\theta)$. Substituting $\iota(x,\theta)\le f_{Q,x}(\theta)$ [cf. Appendix~\ref{app:B}] into the classical derivations gives Eqs.~\eqref{eq:THM3a} and \eqref{eq:THM3b}. In addition to the conditions of Remarks~1 and~2, saturation requires equality in both quantum inequalities: $\operatorname{Im}\Tr(\Pi_xL_\theta\rho_\theta)=0$ and proportionality of $\sqrt{\rho_\theta}\sqrt{\Pi_x}$ and $\sqrt{\rho_\theta}L_\theta\sqrt{\Pi_x}$.

\subsection{Status of the interference term}
\label{sec:interference_status}

Because the CQFI is outcome dependent, the quantum bound fluctuates from one result to another. Separately, the decomposition \eqref{eq:decomp} resolves the projected SLD second moment into physically distinct channels. It is tempting to read a negative cross-term operationally, as a signal that the measurement basis is poorly aligned with the parameter-encoding direction.

We have tested this reading in the model of Sec.~\ref{sec:qubit} and it does not survive. Two obstructions are worth recording, since both are structural rather than particular to that model.

First, neither the CQFI of one outcome nor its probability-weighted mean selects a POVM. Indeed,
\begin{equation}
\sum_xp(x|\theta)f_{Q,x}(\theta)=\mathcal F_Q(\theta)
\end{equation}
for every complete POVM, so this average is measurement independent. A basis-dependent local objective is instead the classical FI of the selected measurement,
\begin{equation}
F_{\{\Pi_x\}}(\theta)=\sum_xp(x|\theta)\iota(x,\theta),
\qquad \max_{\{\Pi_x\}}F_{\{\Pi_x\}}(\theta)=\mathcal F_Q(\theta).
\end{equation}
For a finite prior, expected information gain provides another measurement-dependent objective.

Second, the sign of the cross-term is not a monotone of measurement quality. In the thermal qubit it depends on the azimuthal angle through $\cos\phi$, Eq.~\eqref{eq:fX_qubit_explicit}, whereas the classical FI is built from the score of the complete outcome distribution. No ordering theorem relates these quantities, and configurations with a negative cross-term need not have smaller classical FI than their sign-reversed counterparts.

We therefore make no operational claim for the interference term. The decomposition characterizes how population and coherent channels combine in the projected SLD second moment; it does not optimize the measurement entering the PMI bound. Whether another outcome-dependent functional can serve that purpose remains open.
\endgroup

\section{EXAMPLES}
\label{sec:examples}

{\color{changecol}To illustrate the bounds we consider two systems: a classical stochastic process, in which elapsed time is inferred from the instantaneous position of a Brownian particle, and a quantum system, in which a transverse field is estimated with a thermal qubit. The second example activates all three terms of the projected-SLD decomposition \eqref{eq:decomp}.}

\subsection{Overdamped Langevin dynamics}
\label{sec:langevin}

\begin{figure}[ht!]
    \centering
\includegraphics[width=1.0\linewidth]{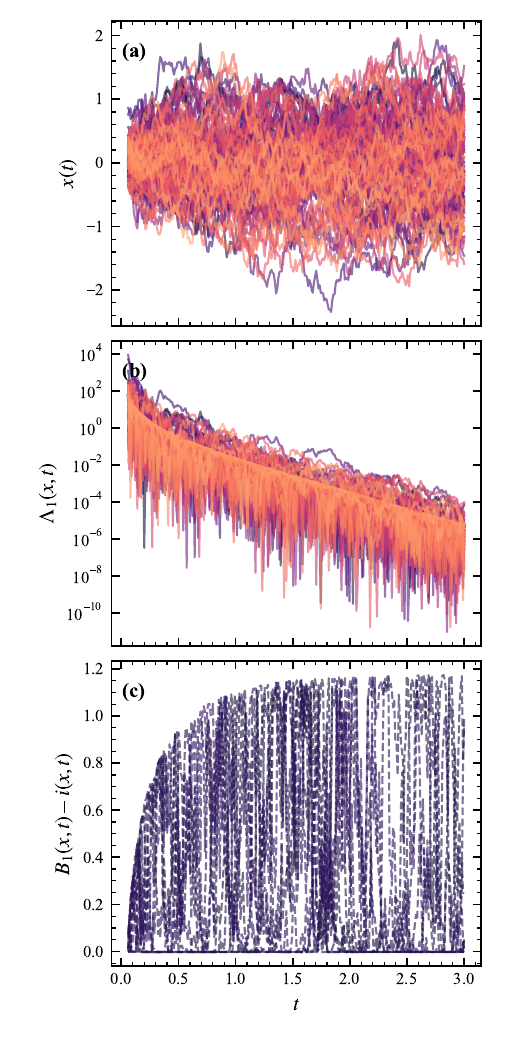}
    \caption{Application of Theorem~1 to overdamped Brownian motion. (a) Evolution of $N_{{\rm traj}}=100$ single-shot trajectories $x(t)$. (b) Stochastic Fisher information $\Lambda_1(x,t)=\iota(x,t)$ along each trajectory. (c) Dynamic pointwise-information gap for a representative subset of trajectories across the expanding observation window. We plot $B_1(x,t)-i(x,t)$, the difference between the right- and left-hand sides of Eq.~\eqref{eq:THM1}; non-negativity at all times verifies the bound.}
    \label{fig:ex1}
\end{figure}

Consider a Brownian particle trapped in a one-dimensional harmonic potential, a standard model in stochastic thermodynamics. The system is governed by the overdamped Langevin equation
\begin{equation}
    \dot{x} = -k x + \sqrt{2D}\xi(t)
    \label{eq:langevin}
\end{equation}
\begingroup
\color{changecol}
where $k$ is the trap stiffness, $D$ is the diffusion coefficient, and $\xi(t)$ is white noise. We take the elapsed time $t$ as the unknown parameter and the instantaneous position $x(t)$ as the observed outcome. The simulated paths provide a sequence of such readouts, but the likelihood below is the single-time position density, not the density of the complete path. Prior to stationarity it is Gaussian,
\begin{equation}
    p(x|t) = \frac{1}{\sqrt{2\pi \sigma^2(t)}} \exp\left(-\frac{x^2}{2\sigma^2(t)}\right),
    \label{eq:gaussian}
\end{equation}
with variance $\sigma^2(t) = \frac{D}{k}(1 - e^{-2kt})$.

For a window $[t_0,t_f]$, Theorem~1 applies with a uniform time prior. We denote the corresponding prior-predictive position density by
\begin{equation}
    p_{t_f}(x) = \frac{1}{t_f - t_0}\int_{t_0}^{t_f} dt'\, p(x|t').
    \label{eq:marginal_langevin}
\end{equation}

The SFI with respect to time, which is directly the functional $\Lambda_1(x,t)$, follows from Eqs.~\eqref{eq:SFI} and \eqref{eq:gaussian},
\begin{align}
    \Lambda_1(x,t) = \iota(x,t)  = \left[ \frac{\partial_t\sigma^2(t)}{2\sigma^2(t)} \left( \frac{x^2}{\sigma^2(t)} - 1 \right) \right]^2
    \label{eq:SFI_langevin}
\end{align}
where $\partial_t\sigma^2(t)=2D e^{-2kt}$.

Theorem 1 then bounds the PMI by $i(x,t) \le B_1(x,t)$, with the bound given explicitly by
\begin{align}
    B_1(x,t) = \log\left[ \frac{p(x|t_0) + p(x|t)}{2p_t(x)} \right. \nonumber \\ \left. + \frac{1}{2}\int_{t_0}^{t} dt' \, \frac{p(x|t')}{p_t(x)} \sqrt{\iota(x,t')} \right],
    \label{eq:B1_explicit}
\end{align}
where the expanding-window marginal $p_t(x)$ is Eq.~\eqref{eq:marginal_langevin} with $t_f=t$. The PMI is correspondingly $i(x,t)=\log[p(x|t)/p_t(x)]$.

{\color{changecol}We evaluate the integral in Eq.~\eqref{eq:B1_explicit} numerically because $\sqrt{\iota(x,t')}$ contains $|x^2/\sigma^2(t')-1|$ and therefore has a realization-dependent kink when $x^2=\sigma^2(t')$. We use the composite trapezoidal rule on the same expanding uniform grid as the Euler--Maruyama simulation, with $k=1$, $D=0.5$, $N_{\rm traj}=100$, $\Delta t=10^{-2}$, $t_0=0.05$, and $T=3$. The initial positions are sampled from the Gaussian \eqref{eq:gaussian} at $t_0$. At time $t_j$, the marginal and the bound integral use the $j+1$ nodes $t_0,\ldots,t_j$, up to 296 nodes at $T$. No adaptive quadrature tolerance is involved\footnote{The code used to analyze the data is available from Ref. \cite{pbmelo_repository}}.}
\endgroup

Figure \ref{fig:ex1} shows the validation of Eq.~\eqref{eq:THM1}. Since the conditional distribution \eqref{eq:gaussian} is known in closed form, both $i(x,t)$ and $B_1(x,t)$ can be computed for every realization, and the bound tested directly. For all generated trajectories the bound is respected at all times. Rare, large excursions from the origin produce sharp spikes in $\Lambda_1(x,t)$ [Fig.~\ref{fig:ex1}(b)], illustrating how individual realizations can carry far more sensitivity to the elapsed time than the ensemble average would suggest.

Panel (c) requires a comment on how the bound is displayed. Unlike a deterministic bound, $B_1(x,t)$ is itself a random variable: it depends on the realized position $x(t)$ through the conditional probability $p(x|t')$, through the SFI $\iota(x,t')$, and through the marginal $p(x)$. There is consequently no single fixed ceiling that can be overlaid on the family of curves $i(x,t)$ --- each trajectory carries its own bound, and a plot of both quantities would show as many ceilings as realizations. Plotting instead the difference $B_1(x,t) - i(x,t)$ collapses all realizations onto a common axis, on which the statement of Theorem 1 is simply the positivity of every curve. This is the generic situation for stochastic bounds, and distinguishes them from their ensemble counterparts, for which a single deterministic ceiling does exist.

\begingroup
\color{changecol}
\subsection{Field sensing with a thermal qubit}
\label{sec:qubit}

We now turn to a quantum example. We consider a two-level system used to estimate a transverse field $\theta$, following the model of Ref.~\cite{melo2026}, with Hamiltonian
\begin{equation}
    \hat{H}_S(\theta) = \Delta \sigma_z + \frac{\theta}{2}\sigma_x,
    \label{eq:qubit_H}
\end{equation}
where $2\Delta$ is the zero-field energy gap. The system is prepared in the thermal state $\rho_\theta = e^{-\beta \hat{H}_S(\theta)}/\mathcal{Z}(\theta)$ with $\mathcal{Z}(\theta) = \text{Tr}(e^{-\beta \hat{H}_S(\theta)})$, and the field lies in a finite interval $\theta \in [a,b]$ with uniform prior, so that Theorem~3 applies in the form \eqref{eq:THM3a}.

\begingroup
\color{changecol}
We choose this model in preference to the more familiar pure-state phase estimation protocol for a specific reason. For a pure state undergoing unitary evolution, $\rho_\theta = |\psi(\theta)\rangle\langle\psi(\theta)|$, the eigenvalues are $\{1,0\}$ independently of $\theta$, so $\partial_\theta p_n = 0$ and both the incoherent and the cross-term of Eq.~\eqref{eq:decomp} vanish identically: $\tilde{f}^{IC}_{Q} = \tilde{f}^{X}_{Q} = 0$. The projected SLD second moment is then purely coherent, but the true CQFI $f_{Q,x}$ can still depend on the outcome and need not equal $\tilde f_{Q,x}$ outcome by outcome. The thermal qubit \eqref{eq:qubit_H}, by contrast, has $\theta$-dependent populations \textit{and} a $\theta$-dependent eigenbasis, and therefore activates all three projected-SLD channels.

For a finite-prior likelihood the readout must be one fixed experiment: the POVM cannot depend on the unknown integration variable. We therefore use laboratory-frame projectors
\begin{equation}
\Pi_s=\frac{1}{2}\left(\mathbf 1+s\,\bm m\cdot\bm\sigma\right),
\qquad s=\pm1,
\label{eq:fixed_qubit_POVM}
\end{equation}
with $\bm m$ independent of $\theta$. The numerical curves are obtained by a deterministic sweep of the exact $2\times2$ Gibbs state and its SLD for the two possible outcomes; no Monte Carlo sampling is used. The physical readout remains stochastic because a single run returns either $s=+1$ or $s=-1$ with Born probability $p_s(\theta)=\Tr(\Pi_s\rho_\theta)$.
\endgroup

\subsubsection{Spectrum and populations}

The eigenvalues of \eqref{eq:qubit_H} are $E_\pm(\theta) = \pm\Omega(\theta)/2$ with generalized Rabi frequency
\begin{equation}
    \Omega(\theta) = \sqrt{4\Delta^2 + \theta^2}.
    \label{eq:rabi}
\end{equation}
We note that Ref.~\cite{melo2026} quotes $\Omega=\sqrt{\Delta^2+\theta^2}$ for the same Hamiltonian; the factor of four follows from $\hat{H}_S = \Delta\sigma_z + \tfrac{\theta}{2}\sigma_x$ having eigenvalues $\pm\sqrt{\Delta^2+\theta^2/4} = \pm\tfrac{1}{2}\sqrt{4\Delta^2+\theta^2}$, and we have verified Eq.~\eqref{eq:rabi} against direct numerical diagonalization. The equilibrium populations are
\begin{equation}
    p_\pm(\theta) = \frac{1}{2}\left[1 \mp \tanh\left(\frac{\beta\Omega(\theta)}{2}\right)\right],
    \label{eq:populations}
\end{equation}
and their logarithmic derivatives, which constitute the classical population channel, are
\begin{equation}
    \frac{\partial_\theta p_\pm}{p_\pm} = -\frac{\beta\theta}{2\Omega(\theta)}\left[\tanh\left(\frac{\beta\Omega(\theta)}{2}\right) \pm 1\right].
    \label{eq:logderiv_pop}
\end{equation}
Their sum, which will control the interference term, is
\begin{equation}
    \frac{\partial_\theta p_+}{p_+} + \frac{\partial_\theta p_-}{p_-} = -\frac{\beta\theta}{\Omega(\theta)}\tanh\left(\frac{\beta\Omega(\theta)}{2}\right),
    \label{eq:sum_logderiv}
\end{equation}
which is strictly negative for $\theta>0$ and $\beta>0$. Choosing a real, smooth eigenvector gauge and writing the eigenbasis in terms of a mixing angle $\varphi(\theta)$, the eigenvectors rotate in the $x$-$z$ plane at rate
\begin{equation}
    \langle +(\theta)|\partial_\theta -(\theta)\rangle = \frac{\Delta}{\Omega^2(\theta)},
    \label{eq:mixing}
\end{equation}
which is the geometric factor controlling the coherent and interference channels, and which we have likewise verified numerically.

\subsubsection{Local decomposition in an energy-basis direction}

It is useful first to inspect the decomposition at a reference field $\theta_0$. We freeze the projectors $\Pi_\pm=|\pm(\theta_0)\rangle\langle\pm(\theta_0)|$ and evaluate the local contributions at $\theta=\theta_0$. At that point the outcome probabilities equal the thermal populations $p_\pm(\theta_0)$, while away from $\theta_0$ the likelihood is correctly given by $\Tr[\Pi_\pm\rho_\theta]$ rather than by $p_\pm(\theta)$. The following expressions are therefore local identities at the reference field, not a parameter-dependent measurement protocol over the prior.
The incoherent and coherent contributions follow from Eqs.~\eqref{eq:B_fIC} and \eqref{eq:B_fC} with $c_n = \delta_{n,\pm}$,
\begin{align}
    \tilde{f}^{IC}_{Q,\Pi_\pm}(\theta_0) &= \frac{\beta^2\theta_0^2}{4\Omega^2(\theta_0)}\left[\tanh\left(\frac{\beta\Omega(\theta_0)}{2}\right)\pm 1\right]^2,
    \label{eq:fIC_qubit} \\
    \tilde{f}^{C}_{Q,\Pi_\pm}(\theta_0) &= \frac{4\Delta^2}{\Omega^4(\theta_0)}\tanh^2\left(\frac{\beta\Omega(\theta_0)}{2}\right).
    \label{eq:fC_qubit}
\end{align}
Equation \eqref{eq:fIC_qubit} is the square of Eq.~\eqref{eq:logderiv_pop} and encodes information gained from thermal population shifts; Eq.~\eqref{eq:fC_qubit} is controlled by the mixing rate \eqref{eq:mixing} and encodes information gained from the rotation of the energy eigenbasis. The coherent term vanishes at infinite temperature ($\beta=0$) or in the absence of a gap ($\Delta=0$), while the incoherent term vanishes at $\theta=0$, where the field does not yet split the levels.

For this measurement the cross-term is absent at the reference field. With $|\alpha\rangle=|\pm(\theta_0)\rangle$, $c_n=\delta_{n,\pm}$, and since the sum in Eq.~\eqref{eq:B_fX} runs strictly over $k\neq n$, every contribution is multiplied by $c_k^*c_n=0$. Hence
\begin{equation}
    \tilde{f}^{X}_{Q,\Pi_\pm}(\theta_0) = 0, \qquad \tilde{f}_{Q,\Pi_\pm}(\theta_0) = \tilde{f}^{IC}_{Q,\Pi_\pm}(\theta_0) + \tilde{f}^{C}_{Q,\Pi_\pm}(\theta_0).
    \label{eq:fX_eigen}
\end{equation}
The finite-prior PMI is instead constructed from the fixed likelihood
\begin{align}
&p(s|\theta)=\Tr(\Pi_s\rho_\theta),\qquad
\nonumber p(s)=\frac{1}{b-a}\int_a^bdu\,p(s|u), \\& 
i(s,\theta)=\log\frac{p(s|\theta)}{p(s)}.
\label{eq:PMI_qubit_fixed}
\end{align}
Theorem~3 then gives
\begin{align}
    i(s,\theta) \le \log\left[\frac{p(s|a) + p(s|b)}{2p(s)} \right. \nonumber \\ \left. + \frac{1}{2p(s)}\int_a^b du\, p(s|u)\sqrt{f_{Q,s}(u)}\right],
    \label{eq:bound_qubit_fixed}
\end{align}
with $f_{Q,s}$ from Eq.~\eqref{eq:CQFI}; the remaining integral is one-dimensional and evaluated by quadrature.

\subsubsection{Measurement outside the eigenbasis: the interference term}

The cross-term is activated by measuring along a direction that is not an eigenstate of $\rho_\theta$. For a local analytic diagnosis at a reference field $\theta_0$, consider the fixed rank-one projector $\Pi_\alpha = |\alpha\rangle\langle\alpha|$ with
\begin{equation}
    |\alpha\rangle = \cos\frac{\chi}{2}|+(\theta_0)\rangle + e^{i\phi}\sin\frac{\chi}{2}|-(\theta_0)\rangle,
    \label{eq:probe}
\end{equation}
so that both $c_\pm$ are non-zero, with $\text{Re}(c_+^* c_-) = \tfrac{1}{2}\cos\phi\sin\chi$. For a two-level system the sum in Eq.~\eqref{eq:B_fX} contains the single pair $(k,n)=(+,-)$ and its conjugate. Using $p_+ + p_- = 1$, $p_- - p_+ = \tanh(\beta\Omega/2)$, together with Eqs.~\eqref{eq:sum_logderiv} and \eqref{eq:mixing}, it collapses to the closed form
\begin{equation}
    \tilde{f}^{X}_{Q,\alpha}(\theta) = -2\cos\phi\,\sin\chi\;\frac{\beta\,\theta\,\Delta}{\Omega^3(\theta)}\,\tanh^2\!\left(\frac{\beta\Omega(\theta)}{2}\right).
    \label{eq:fX_qubit_explicit}
\end{equation}
Equation~\eqref{eq:fX_qubit_explicit} is understood locally, at the field where the probe is constructed (set $\theta=\theta_0$). We have verified it against direct evaluation of Eq.~\eqref{eq:B_fX} from numerically obtained spectra, finding agreement to machine precision across randomly sampled $(\Delta,\beta,\theta_0,\chi,\phi)$. For a sweep over a finite prior, the fixed laboratory projector \eqref{eq:fixed_qubit_POVM} is used instead and every quantity is evaluated directly from the matrices.

Equation \eqref{eq:fX_qubit_explicit} makes the mechanism transparent. The sign of the cross-term is set by $\cos\phi$, that is by the azimuthal orientation of the probe relative to the plane in which the eigenbasis rotates. Probes with $\cos\phi>0$ and $\sin\chi>0$ give $\tilde{f}^{X}_{Q,\alpha}<0$ for $\theta>0$: the population shift \eqref{eq:sum_logderiv} and the geometric rotation \eqref{eq:mixing} act in opposition relative to $|\alpha\rangle$, and the trajectory-level sensitivity is reduced below what the two channels would supply independently. The magnitude is maximal at $\chi=\pi/2$, $\phi\in\{0,\pi\}$ --- an equal superposition of the two eigenstates, lying in the rotation plane --- and the term vanishes in four distinct limits: at $\chi=0,\pi$, recovering Eq.~\eqref{eq:fX_eigen}; at $\phi=\pm\pi/2$, where the probe is orthogonal to the rotation plane; at $\theta=0$, where the populations are stationary; and at $\beta\to 0$, where the thermal state carries no information at all. Each limit has a transparent physical reading, and together they confirm that the interference channel requires \textit{both} a moving population and a rotating basis, sampled by a probe sensitive to both.

Two caveats must be stated. First, the criterion \eqref{eq:chain_cond} is not satisfied for every tilted probe in this model: with $\langle\alpha|\rho_\theta|\alpha\rangle=p_+\cos^2(\chi/2)+p_-\sin^2(\chi/2)$, the weighted sum in Eq.~\eqref{eq:chain_cond} can take either sign as the probe is varied. Theorem~3 is controlled by $f_{Q,\alpha}$ throughout; Eq.~\eqref{eq:fX_qubit_explicit} characterizes the interference structure of the projected SLD second moment, not the numerical value of the bound.

Second, the sign of $\tilde{f}^{X}_{Q,\alpha}$ should not be read as a verdict on the measurement. The cross-term is antisymmetric under $\phi\to\phi+\pi$, whereas the classical FI is constructed from the complete two-outcome score and is not ordered by this sign. Thus destructive interference in Eq.~\eqref{eq:fX_qubit_explicit} describes how population and coherence channels combine in the projected SLD second moment; it is not a diagnosis of a poorly chosen basis.

Figure~\ref{fig:thermal_fixed_readout} evaluates the finite-prior construction with $\Delta=1$, $\beta\Delta=1.5$, $\theta/\Delta\in[-2,2]$, and a fixed laboratory direction $\bm m=(\sin\eta,0,\cos\eta)$ with $\eta=0.68\pi$. The field grid contains 2401 points. At each point we construct the exact Gibbs state, its analytic derivative, and the SLD; the prior marginal and the integral in Eq.~\eqref{eq:bound_qubit_fixed} are then evaluated by trapezoidal quadrature. Direct Hamiltonian diagonalization, centered finite differences, the Lyapunov equation, $\sum_s p_s f_{Q,s}=\mathcal F_Q$, and $\iota_s\le f_{Q,s}$ provide independent numerical checks. The calculation is deterministic; the outcome label indexes the stochastic result that would occur in one physical readout.

\begin{figure}[ht!]
\centering
\includegraphics[width=0.9\columnwidth]{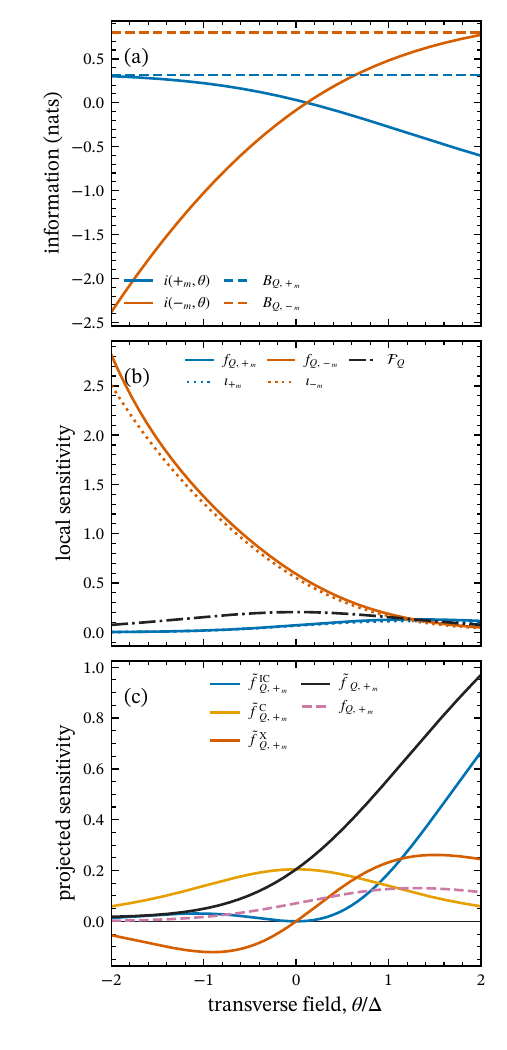}
\caption{\label{fig:thermal_fixed_readout}
Outcome-resolved thermal-qubit field sensing with the fixed laboratory measurement as in Eq.~\eqref{eq:fixed_qubit_POVM}.
(a) PMI for the two outcomes and their finite-support quantum bounds.
(b) SFI $\iota_s$ and CQFI $f_{Q,s}$ for each outcome, together with the QFI $\mathcal F_Q$.
(c) Incoherent, coherent, and cross contributions to the projected SLD second moment for outcome $+_{\bm m}$, their sum, and the true CQFI. The cross contribution can be negative while the projected second moment and CQFI remain non-negative.}
\end{figure}
\endgroup

\section{DISCUSSION AND CONCLUSIONS}
\label{sec:conclusions}

{\color{changecol}We have established two upper bounds on the pointwise mutual information in terms of the stochastic Fisher information: one for a bounded parameter interval and one for regular differentiable priors with the required boundary behavior. The central question is whether the information density assigned to one realized outcome can be constrained by the squared score of that same outcome before averaging. The answer is affirmative: a total-variation argument gives the pointwise bounds and their saturation conditions. Averaging the pointwise inequality and applying Jensen's inequality twice yields the ensemble MI--FI bound of Ref.~\cite{Gorecki_2025}. The latter is therefore a corollary of the pointwise result and two additional relaxations, not literally the probability average of the pointwise right-hand side.}

{\color{changecol}For a fixed, parameter-independent quantum measurement, Theorem~3 follows from $\iota(x,\theta)\le f_{Q,x}(\theta)$. The CQFI is attached to the realized classical outcome and averages to the QFI by completeness. The Langevin example evaluates the classical bound for instantaneous position readouts along simulated paths. The thermal-qubit example evaluates the finite-prior quantum bound from exact two-dimensional matrices and, separately, gives analytic expressions for the population, coherent, and interference contributions to the projected SLD second moment.}

{\color{changecol}This separation is essential. The bound is controlled by $f_{Q,x}$, whereas the three-channel decomposition belongs to $\tilde f_{Q,x}$. Neither $f_{Q,x}\le\tilde f_{Q,x}$ nor the reverse holds universally. Moreover, $\sum_xp(x|\theta)f_{Q,x}(\theta)=\mathcal F_Q(\theta)$ for every complete POVM, so neither the CQFI nor its mean selects a measurement basis. Basis optimization instead uses a measurement-dependent objective such as $F_{\{\Pi_x\}}=\sum_xp(x|\theta)\iota(x,\theta)$ or Bayesian expected information gain. The sign of the projected-SLD cross-term has no established operational role in this optimization. Its unweighted trace vanishes, whereas its probability-weighted mean generally does not. We therefore make no adaptive-metrology claim based on that sign.}

{\color{changecol}The saturation conditions are stringent, and evaluating the outcome-resolved quantities still requires a calibrated likelihood or state model. Natural extensions include multiparameter bounds, single-record feedback in stochastic thermodynamics, and continuously monitored quantum systems. Syndrome-based noise tracking is a timely but strictly prospective application: a measured syndrome is a classical outcome, so Theorems~1 and~2 apply once a policy-conditioned likelihood $p(s|\theta)$ is available, while Theorem~3 supplies the corresponding quantum constraint for a fixed syndrome-extraction measurement. Real QEC records contain measurement faults, leakage, and space--time correlations and are not generally binomial; constructing and validating the required circuit-level likelihood is left for future work \cite{Wagner2021,Wagner2022,Sivak2026}.}

{\color{changecol}On experimental accessibility, evaluating either quantum object requires the SLD $L_\theta$, so neither bypasses characterization of the dynamics. The projected SLD second moment avoids reconstructing the full spectrum of $\rho_\theta$ once the action of $L_\theta$ along the measured direction is available. Modulated-Hamiltonian and time-dependent-response protocols provide possible routes to this information without full tomography \cite{hauke2016measuring,yu2022quantum}.}

\section*{Acknowledgments.} We acknowledge S\'ilvio Queir\'os, Diogo Soares-Pinto, Luca Innocenti, Francesco Albarelli and Welles Morgado for discussions, Pedro Paraguass\'u and Alessandro Candeloro for reading of the manuscript. Special thanks to Diogo Oliveira Soares-Pinto for drawing my attention to the motivations to this problem. This work is supported by the Brazilian agencies CAPES, finance code 001, and CNPq, grant no. 140264/2026-4. 
\section*{AI-Assisted research statement}
ChatGPT has been useful for this work. I acknowledge the use of ChatGPT 5.6 for checking the proof of the derived Theorems in this paper after receiving the first referee comments. I also acknowledge the use of ChatGPT 5.6 for assistance in grammar and cohesion matters in the manuscript's text. Although AI has been used in this work, the author takes full responsibility for its content.
\providecommand{\newblock}{}
\bibliography{refs}

\onecolumngrid

\appendix

\section{Proof of PMI bounds \label{app:A}}
We begin by expressing the PMI using an arbitrary non-negative function $f(\theta)$ such that $\text{supp } f(\theta) \supseteq \text{supp } p(\theta)$. The PMI can be written algebraically as
\begin{equation}
    i(x,\theta) = \log\frac{p(x|\theta)f(\theta)}{p(x)} - \log f(\theta).
    \label{eq:A1}
\end{equation}

{\color{changecol}To bound the first term, we assume that $p(x|\theta)f(\theta)\to0$ as $\theta\to\pm\infty$ and that this product has finite total variation. Its value at a specific parameter is bounded by its global maximum, which is bounded in turn by half the integral of its absolute derivative:}
\begin{equation}
    \frac{p(x|\theta)f(\theta)}{p(x)} \le \max_{\theta'} \left( \frac{p(x|\theta')f(\theta')}{p(x)} \right) \le \frac{1}{2} \int_{-\infty}^{\infty} d\theta' \left| \frac{d}{d\theta'} \left( \frac{p(x|\theta')f(\theta')}{p(x)} \right) \right|.
    \label{eq:A2}
\end{equation}

We now expand the derivative inside the absolute value using the product rule. Recognizing that probability distributions are strictly non-negative, we can factor out $\frac{p(x|\theta')}{p(x)}$. Utilizing the identity $\frac{\partial p(x|\theta')}{\partial \theta'} = p(x|\theta')\partial_{\theta'}\log p(x|\theta')$ and the definition of the stochastic Fisher information (SFI) $\iota(x,\theta') = (\partial_{\theta'}\log p(x|\theta'))^2$, the exact absolute variation is determined by the square root of a functional 
\begin{equation}
    \left| \frac{d}{d\theta'} \left( \frac{p(x|\theta')f(\theta')}{p(x)} \right) \right| = \left| \frac{p(x|\theta')}{p(x)} \right| \sqrt{\Lambda(x,\theta')},
\end{equation}
where 
\begin{equation}
    \Lambda(x,\theta) \equiv \iota(x,\theta)f(\theta)^2 + \dot{f}(\theta)^2 + 2\partial_\theta \log p(x|\theta) f(\theta)\dot{f}(\theta).
    \label{eq:A4}
\end{equation}

Substituting this back into our original PMI expression yields the generalized bound:
\begin{equation}
    i(x,\theta) \le \log \left( \frac{1}{2} \int_{-\infty}^{\infty} d\theta' \left|\frac{p(x|\theta')}{p(x)}\right| \sqrt{\Lambda(x,\theta')} \right) - \log f(\theta).
    \label{eq:A5}
\end{equation}

\subsection{Proof of Theorem 1}

{\color{changecol}If $\operatorname{supp}p(\theta)\subseteq[a,b]$, we choose $f(\theta)$ to be the indicator of $[a,b]$.}

To evaluate the generalized bound without squaring the derivative of a step function, we apply the triangle inequality $|A+B| \le |A| + |B|$ directly to the expanded absolute derivative prior to integration. This subadditivity allows us to decouple the stochastic Fisher information from the boundary evaluations:
\begin{equation}
    \left| \frac{\partial p(x|\theta')}{\partial \theta'} f(\theta') + p(x|\theta')\dot{f}(\theta') \right| \le p(x|\theta')\sqrt{\iota(x,\theta')}f(\theta') + p(x|\theta')|\dot{f}(\theta')|.
\end{equation}

The absolute derivative of the step function yields Dirac delta functions at the boundaries, $|\dot{f}(\theta')| = \delta(\theta'-a) + \delta(\theta'-b)$. When integrating this absolute variation over all space and applying the $1/2$ factor, these delta functions extract the exact boundary probability conditions $\frac{p(x|a)}{2p(x)} + \frac{p(x|b)}{2p(x)}$. 

Within the open interval $(a, b)$, the derivative $\dot{f}(\theta')$ vanishes, and $f(\theta')=1$. Evaluating the functional in this region isolates the stochastic Fisher information, yielding 
\begin{equation}
    \Lambda_1(x,\theta) = \iota(x,\theta).
    \label{eq:A9}
\end{equation}

Integrating the absolute variation across the entire domain, including the boundaries at $a$ and $b$, we recover the properly bounded terms. Finally, because $f(\theta) = 1$ for $\theta \in [a,b]$, the penalty term $-\log f(\theta)$ vanishes. This yields
\begin{equation}
    i(x,\theta) \le \log\left(\frac{p(x|a)+p(x|b)}{2p(x)} + \frac{1}{2}\int_{a}^{b}d\theta\left|\frac{p(x|\theta)}{p(x)}\right|\sqrt{\Lambda_{1}(x,\theta)}\right).
    \label{eq:A10}
\end{equation}

\subsection{Proof of Theorem 2}

Theorem 2 follows directly from the generalized bound by choosing the arbitrary function to be the exact prior probability distribution, $f(\theta) = p(\theta)$. Substituting $f(\theta) = p(\theta)$ into our definition of $\Lambda(x,\theta)$ immediately recovers 
\begin{equation}
    \Lambda_2(x,\theta) = \iota(x,\theta)p(\theta)^2 + \dot{p}(\theta)^2 + 2\partial_\theta \log p(x|\theta)\dot{p}(\theta)p(\theta).
    \label{eq:A7}
\end{equation}

{\color{changecol}The second term, $-\log f(\theta)$, becomes the prior surprisal $-\log p(\theta)=s(\theta)$. Under the regularity and boundary assumptions stated in Theorem~2, the generalized bound becomes}
\begin{equation}
    i(x,\theta) \le \log \left( \frac{1}{2} \int_{-\infty}^{\infty} d\theta \left|\frac{p(x|\theta)}{p(x)}\right| \sqrt{\Lambda_2(x,\theta)} \right) + s(\theta).
    \label{eq:A8}
\end{equation}

\section{The Conditional QFI, the projected SLD second moment, and the spectral decomposition \label{app:B}}

\begingroup
\color{changecol}
Following Ref.~\cite{melo2026}, we derive the CQFI for a general fixed POVM with positive effects $\Pi_x\ge0$ and $\sum_x\Pi_x=\mathbf1$. The later projected-SLD decomposition is restricted to rank-one projectors $\Pi_\alpha=|\alpha\rangle\langle\alpha|$. The POVM is independent of $\theta$ throughout, so $\partial_\theta p(x|\theta)=\Tr(\Pi_x\partial_\theta\rho_\theta)$.

\subsection{From the surprisal rate to the CQFI}

For an unknown parameter $\theta$, the classical SFI associated with a single measurement outcome $x$ is
\begin{equation}
    \iota(\theta,x) = \left(\frac{\partial \log p(x|\theta)}{\partial \theta}\right)^2.
    \label{eq:B_SFI}
\end{equation}
In the quantum setting the conditional probability is fixed by the Born rule, $p(x|\theta) = \mathrm{Tr}(\Pi_x \rho_\theta)$. The standard QFI is the maximum of the classical Fisher information over all POVMs,
\begin{equation}
    \mathcal{F}_{Q}(\theta) = \mathrm{Tr}\left(\rho_{\theta}L_{\theta}^2\right) = \max_{\{\Pi_x\}} \mathcal{I}(\theta),
    \label{eq:B_QFI}
\end{equation}
where $L_\theta$ is the Hermitian solution of the Lyapunov equation $\partial_\theta \rho_\theta = \frac{1}{2}\{L_\theta,\rho_\theta\}$. Since $\rho_\theta$ and $L_\theta$ are Hermitian, cyclicity gives $\mathrm{Tr}(\rho_\theta L_\theta^2) = \mathrm{Tr}(L_\theta \rho_\theta L_\theta)$, an identity used repeatedly below.

Expanding the logarithmic derivative in Eq.~\eqref{eq:B_SFI} and inserting the Born rule,
\begin{equation}
    \iota(\theta,x) = \left(\frac{1}{p(x|\theta)}\frac{\partial p(x|\theta)}{\partial \theta}\right)^2 = \left(\frac{\mathrm{Tr}\left(\Pi_{x} \partial_\theta \rho_\theta \right)}{\mathrm{Tr}(\Pi_{x}\rho_\theta)}\right)^2.
    \label{eq:B_exact}
\end{equation}
Substituting the SLD identity, the numerator becomes
\begin{equation}
    \mathrm{Tr}(\Pi_x \partial_\theta \rho_\theta) = \tfrac{1}{2}\mathrm{Tr}\left(\Pi_x\{L_\theta,\rho_\theta\}\right) = \mathrm{Re}\,\mathrm{Tr}(\Pi_x L_\theta \rho_\theta),
    \label{eq:B_numerator}
\end{equation}
so that Eq.~\eqref{eq:B_exact} is exactly
\begin{equation}
    \iota(\theta,x) = \frac{\left[\mathrm{Re}\,\mathrm{Tr}(\Pi_x L_\theta \rho_\theta)\right]^2}{\left[\mathrm{Tr}(\Pi_x \rho_\theta)\right]^2}.
    \label{eq:B_exact2}
\end{equation}
To obtain a quantity depending only on the local SLD structure, and recovering the QFI on averaging, we bound Eq.~\eqref{eq:B_exact2} in two steps. First, $[\mathrm{Re}\, z]^2 \le |z|^2$ gives
\begin{equation}
    \iota(\theta,x) \le \frac{\left|\mathrm{Tr}(\Pi_x L_\theta \rho_\theta)\right|^2}{\left[\mathrm{Tr}(\Pi_x \rho_\theta)\right]^2}.
    \label{eq:B_step1}
\end{equation}
Second, we apply the Cauchy-Schwarz inequality $|\mathrm{Tr}(A^\dagger B)|^2 \le \mathrm{Tr}(A^\dagger A)\mathrm{Tr}(B^\dagger B)$ with the choice
\begin{equation}
    A = \sqrt{\rho_\theta}\sqrt{\Pi_x}, \qquad B = \sqrt{\rho_\theta}\, L_\theta \sqrt{\Pi_x}.
    \label{eq:B_CS_choice}
\end{equation}
Using positivity of $\Pi_x$ and cyclicity,
\begin{align}
    \mathrm{Tr}(A^\dagger B) &= \mathrm{Tr}(\Pi_x \rho_\theta L_\theta), \nonumber \\
    \mathrm{Tr}(A^\dagger A) &= \mathrm{Tr}(\Pi_x \rho_\theta) = \mathrm{Tr}(\rho_\theta \Pi_x), \nonumber \\
    \mathrm{Tr}(B^\dagger B) &= \mathrm{Tr}(\Pi_x L_\theta \rho_\theta L_\theta),
    \label{eq:B_CS_terms}
\end{align}
Because $\mathrm{Tr}(\Pi_x\rho_\theta L_\theta)=\mathrm{Tr}(\Pi_xL_\theta\rho_\theta)^*$, the moduli are equal, and
$|\mathrm{Tr}(\Pi_x L_\theta \rho_\theta)|^2 \le \mathrm{Tr}(\rho_\theta \Pi_x)\,\mathrm{Tr}(\Pi_x L_\theta \rho_\theta L_\theta)$. Combining with Eq.~\eqref{eq:B_step1},
\begin{equation}
    \iota(\theta,x) \le \frac{\mathrm{Tr}(\Pi_x L_\theta \rho_\theta L_\theta)}{\mathrm{Tr}(\rho_\theta \Pi_x)} \equiv f_{Q,x}(\theta),
    \label{eq:B_CQFI}
\end{equation}
which is Eq.~\eqref{eq:sfi_le_cqfi} of the main text and serves as the definition of the CQFI.

Equality requires equality in both steps: $\operatorname{Im}\mathrm{Tr}(\Pi_xL_\theta\rho_\theta)=0$ and proportionality of the Hilbert--Schmidt vectors $\sqrt{\rho_\theta}\sqrt{\Pi_x}$ and $\sqrt{\rho_\theta}L_\theta\sqrt{\Pi_x}$.

We stress that Eq.~\eqref{eq:B_CQFI} carries the ordering $L_\theta\rho_\theta L_\theta$, not $L_\theta^2\rho_\theta$. Commutation of $\rho_\theta$ and $L_\theta$ is sufficient for the two operator orderings to agree. The former is manifestly real and non-negative, whereas $\mathrm{Tr}(\Pi_xL_\theta^2\rho_\theta)/\mathrm{Tr}(\rho_\theta\Pi_x)$ can be complex. Non-negativity follows with $M=\sqrt{\rho_\theta}L_\theta$: the numerator is $\mathrm{Tr}[(M\sqrt{\Pi_x})^\dagger(M\sqrt{\Pi_x})]\ge0$.

\subsection{Recovery of the QFI on averaging}

Averaging Eq.~\eqref{eq:B_CQFI} over the outcome statistics removes the conditioning,
\begin{align}
    \langle f_Q \rangle &= \sum_{x} p(x|\theta) f_{Q,x}(\theta) = \sum_x \mathrm{Tr}(\Pi_{x}L_{\theta}\rho_\theta L_\theta) \nonumber \\
    &= \mathrm{Tr}\left[\left(\sum_{x}\Pi_{x}\right)L_{\theta}\rho_\theta L_\theta\right] = \mathrm{Tr}(L_\theta \rho_\theta L_\theta) \nonumber \\
    &= \mathrm{Tr}(\rho_\theta L_\theta^2) = \mathcal{F}_{Q}(\theta),
    \label{eq:B_average}
\end{align}
using the completeness relation $\sum_x \Pi_x = \mathbf{1}$ and cyclicity. The CQFI is therefore an unbiased, outcome-resolved decomposition of the QFI.
\endgroup

\begingroup
\color{changecol}
\subsection{The projected SLD second moment and the equality condition}

For a rank-one projector $\Pi_\alpha = |\alpha\rangle\langle\alpha|$ it is convenient to introduce a second, purely state-conditioned object, the \textit{projected SLD second moment}
\begin{equation}
    \tilde{f}_{Q,\alpha}(\theta) = \mathrm{Tr}(\Pi_\alpha L_\theta^2) = \langle \alpha | L_\theta^2 | \alpha \rangle,
    \label{eq:B_proj}
\end{equation}
which measures the local power of the SLD generator along a chosen direction in Hilbert space. Constructing $L_\theta$ still requires knowledge of the state through the Lyapunov equation; what estimating Eq.~\eqref{eq:B_proj} avoids is reconstructing the full spectrum of $\rho_\theta$, since only the action $L_\theta|\alpha\rangle$ along the chosen direction is needed.

The two definitions are distinct in general. Writing $\rho_\theta = \sum_n p_n |n\rangle\langle n|$ with $c_n = \langle n|\alpha\rangle$, and inserting resolutions of the identity in the eigenbasis,
\begin{equation}
    f_{Q,\alpha}(\theta) = \frac{\langle\alpha|L_\theta\rho_\theta L_\theta|\alpha\rangle}{\langle\alpha|\rho_\theta|\alpha\rangle} = \frac{\sum_m p_m \left|\langle\alpha|L_\theta|m\rangle\right|^2}{\sum_n p_n |c_n|^2},
    \label{eq:B_f_spectral}
\end{equation}
whereas
\begin{equation}
    \tilde{f}_{Q,\alpha}(\theta) = \sum_m \left|\langle\alpha|L_\theta|m\rangle\right|^2.
    \label{eq:B_ftilde_spectral}
\end{equation}
Equation~\eqref{eq:B_f_spectral} is a $p_m$-weighted sum divided by the outcome probability, whereas Eq.~\eqref{eq:B_ftilde_spectral} contains the corresponding unweighted sum. Subtracting them shows that they coincide if and only if
\begin{equation}
\sum_m\left[p_m-\langle\alpha|\rho_\theta|\alpha\rangle\right]
\left|\langle\alpha|L_\theta|m\rangle\right|^2=0,
    \label{eq:B_equality}
\end{equation}
which is Eq.~\eqref{eq:equality_cond} of the main text. Requiring every contributing $p_m$ to equal $\langle\alpha|\rho_\theta|\alpha\rangle$ is sufficient but not necessary, because positive and negative terms may cancel in the weighted sum. Another sufficient case is an SLD eigenvector $L_\theta|\alpha\rangle=\lambda_\alpha|\alpha\rangle$, for which both quantities equal $\lambda_\alpha^2$.

Comparing Eqs.~\eqref{eq:B_f_spectral} and \eqref{eq:B_ftilde_spectral} also yields the exact criterion quoted in the main text. Subtracting the two and clearing the positive denominator $\langle\alpha|\rho_\theta|\alpha\rangle = \sum_n p_n|c_n|^2$,
\begin{equation}
f_{Q,\alpha}(\theta) - \tilde{f}_{Q,\alpha}(\theta) = \frac{\sum_m \left(p_m - \langle\alpha|\rho_\theta|\alpha\rangle\right)\left|\langle\alpha|L_\theta|m\rangle\right|^2}{\langle\alpha|\rho_\theta|\alpha\rangle},
\label{eq:B_difference}
\end{equation}
so that $f_{Q,\alpha}\le\tilde{f}_{Q,\alpha}$ if and only if the numerator is non-positive, which is Eq.~\eqref{eq:chain_cond}. This is a weighted condition, not a termwise one: it compares each eigenvalue $p_m$ to the probe overlap $\langle\alpha|\rho_\theta|\alpha\rangle$, but weights the comparison by $|\langle\alpha|L_\theta|m\rangle|^2$. Requiring instead that $p_m \le \langle\alpha|\rho_\theta|\alpha\rangle$ hold termwise for every contributing $m$ would be sufficient but is far too strong: since $\langle\alpha|\rho_\theta|\alpha\rangle$ is a convex combination of the eigenvalues, it cannot exceed $p_{\max}$ unless $|\alpha\rangle$ is the corresponding eigenvector, in which case Eq.~\eqref{eq:B_equality} already applies. The condition \eqref{eq:B_difference} is genuinely two-sided, and we verify below that both signs occur in the thermal qubit of Sec.~\ref{sec:qubit}.

\subsection{Spectral decomposition of the projected SLD second moment}

For a full-rank state, the SLD in the eigenbasis of $\rho_\theta$ reads
\begin{equation}
    L_{\theta} = \sum_{n}\frac{\partial_{\theta}p_{n}}{p_{n}}|n_{\theta}\rangle\langle n_{\theta}| + 2\sum_{n \neq k}\frac{p_k - p_{n}}{p_{n} + p_{k}}\langle n_{\theta}|\partial_{\theta}k_{\theta}\rangle|n_{\theta}\rangle\langle k_{\theta}|.
    \label{eq:B_SLD}
\end{equation}
For rank-deficient states we use the canonical SLD: terms are included only when $p_n+p_k>0$, and the unsupported kernel--kernel block is set to zero. Expressions containing $\partial_\theta p_n/p_n$ are then understood on the support or by a continuous limit.
Decomposing the SLD into a diagonal (incoherent) component $L_{IC}$ and an off-diagonal (coherent) component $L_{C}$, such that $L_\theta = L_{IC} + L_{C}$, the squared operator is $L_\theta^2 = L_{IC}^2 + L_{C}^2 + \{L_{IC}, L_{C}\}$. Evaluating $\tilde{f}_{Q,\alpha}(\theta) = \langle \alpha | L_\theta^2 | \alpha \rangle$ requires accounting for the cross-terms between diagonal and off-diagonal parts. The cross contribution vanishes under the unweighted trace over a complete basis, but is generally non-zero for a specific state $|\alpha\rangle$ and need not vanish under probability weighting. Hence the projected SLD second moment --- and not the CQFI \eqref{eq:B_CQFI} --- splits into three contributions,
\begin{equation}
    \tilde{f}_{Q,\alpha}(\theta) = \tilde{f}_{Q,\alpha}^{IC}(\theta) + \tilde{f}_{Q,\alpha}^{C}(\theta) + \tilde{f}_{Q,\alpha}^{X}(\theta),
    \label{eq:B_decomp}
\end{equation}
which is Eq.~\eqref{eq:decomp} of the main text. The incoherent contribution,
\begin{equation}
    \tilde{f}_{Q,\alpha}^{IC}(\theta) = \sum_n |\langle n|\alpha\rangle|^2 \left(\frac{\partial_\theta p_n}{p_n}\right)^2,
    \label{eq:B_fIC}
\end{equation}
is analogous to the classical SFI and captures sensitivity arising from changes in the eigenvalues. The coherent contribution,
\begin{equation}
    \tilde{f}_{Q,\alpha}^{C}(\theta) = \sum_k \left|\sum_{n (\neq k)} \langle n|\alpha\rangle \frac{2(p_n - p_k)}{p_n + p_k} \langle k|\partial_\theta n\rangle\right|^2,
    \label{eq:B_fC}
\end{equation}
captures information arising from the unitary rotation of the eigenbasis, reflecting the fact that the eigenstates depend on $\theta$ and generally do not commute with their derivatives. Finally, the cross-term,
\begin{equation}
    \tilde{f}_{Q,\alpha}^{X}(\theta) = \sum_{k \neq n} \mathrm{Re}\left[ c_{k}^{*}c_n \left(\frac{\partial_\theta p_k}{p_k} + \frac{\partial_\theta p_n}{p_n}\right) \left(\frac{2(p_n - p_k)}{p_n + p_k}\right) \langle k|\partial_\theta n\rangle \right],
    \label{eq:B_fX}
\end{equation}
with $c_n = \langle n|\alpha\rangle$, quantifies the correlation between the population and coherent channels. Unlike the incoherent and coherent contributions, which are non-negative, $\tilde{f}_Q^X$ can be negative; a negative value indicates destructive interference, in which population shifts and geometric rotations partially cancel relative to $|\alpha\rangle$.

The cross-term obeys an unweighted trace identity. With $\Pi_\alpha = |\alpha\rangle\langle\alpha|$ and $c_n = \langle n|\alpha\rangle$, summing $c_k^* c_n = \langle\alpha|k\rangle\langle n|\alpha\rangle$ over a complete orthonormal basis uses $\sum_\alpha |\alpha\rangle\langle\alpha| = \mathbf{1}$, giving $\sum_\alpha c_k^* c_n = \langle n|k\rangle = \delta_{nk}$. Since Eq.~\eqref{eq:B_fX} runs strictly over $k \neq n$, every contribution is multiplied by $\delta_{nk} = 0$, and
\begin{equation}
    \sum_\alpha\tilde{f}^X_{Q,\alpha}(\theta)
    =\Tr\{L_{IC},L_C\}=0.
    \label{eq:B_fX_avg}
\end{equation}
This is not a stochastic average: in general
$\sum_\alpha p(\alpha|\theta)\tilde{f}^X_{Q,\alpha}(\theta)\neq0$ because
$\sum_\alpha p(\alpha|\theta)|\alpha\rangle\langle\alpha|\neq\mathbf 1$.
The identity that does use probability weights is instead
$\sum_\alpha p(\alpha|\theta)f_{Q,\alpha}(\theta)=\mathcal F_Q(\theta)$.
\endgroup




\end{document}